\documentstyle[epsf,11pt]{article}
\title{A method for solving systems of non-linear
differential equations with moving singularities}
\author{ S. S.
Gousheh$^1$\thanks{Email address: ss-gousheh@cc.sbu.ac.ir}, H. R.
Sepangi$^{1,2}$ and K. Ghafoori-Tabrizi$^1$
\\ {\small\it $^1$Department of physics, Shahid Beheshti
University, Evin, Tehran 19839, Iran}\\ {\small\it $^2$ Institute
for Studies in Theoretical Physics and Mathematics, P.O. Box
19395-5746, Tehran, Iran }}
\begin{document}
\maketitle
\vspace{10mm}
\begin{abstract}
We present a method for solving a class of initial valued,
coupled, non-linear differential equations with `moving
singularities' subject to some subsidiary conditions. We show that
this type of singularities can be adequately treated by
establishing certain `moving' jump conditions across them. We show
how a first integral of the differential equations, if available,
can also be used for checking the accuracy of the numerical
solution.
\end{abstract}
\vspace{5mm}

\hspace{5mm}PACS: 02.60.Cb, 02.60.Lj

\hspace{5mm}{\bf Keywords:} Differential equations, Moving
singularities \vspace{1cm}\\

\section{Introduction}
When solving a physical problem, one usually encounters a set of
coupled nonlinear differential equations, called dynamical
equations or equations of motion, describing the dynamics of the
system. This set of differential equations usually emanate from a
general physical principle, and might include some subsidiary
equations which can be categorized in two distinct classes,
constraints or integrals of motion. The degrees of these
subsidiary equations are usually at least one order lower than the
dynamical equations. At the practical level these two classes of
subsidiary conditions are treated very differently. The first
class has to be solved simultaneously with the rest of the
dynamical equations. However, the integrals of motion need only be
used to put some constraints on the initial conditions. Moreover
they can be used as a consistency check on the solutions.

For physically interesting cases, usually time evolution problems
appear as initial valued ones, and static problems as boundary
valued ones. While solving these problems, one often encounters
various kinds of singularities. These singularities are usually
indications of some profound physical laws or processes with
significant implications. The most common type of singularities
are those in which the coefficient functions of the differential
equations have singularities at some fixed points, for example
$\delta$-functions. These can be called the fixed type of
singularities. In a previous work \cite{gou} we discussed several
general methods to make an efficient numerical algorithm for
boundary valued problems of this type.

In this paper we present a method for handling a class of initial
valued, coupled, non-linear differential equations, whose
solutions contain moving singularities. These singularities have
the property that their positions and severities are {\it apriori}
unknown and depend on the solutions yet to be obtained.
Singularities of this type have been noted in such diverse fields
as celestial mechanics, in particular the classical Kepler problem
\cite{siam}, and in the study of tensor fields defined on moving
surfaces \cite{ut}. We show that this type of singularity can be
adequately treated numerically by establishing certain moving jump
conditions across them. We establish the accuracy of our numerical
solutions by showing that the equation representing the integral
of motion  is satisfied at all values of the independent variable
including at the positions of the moving singularities.

For the integration algorithm we use the basic fourth-order
Runge-Kutta method. It is worth mentioning that more accurate
integration algorithms exist. For example, there are exponential
and Bessel fitted variable step method of order 6 due to Raptis
and Cash \cite{raptis}, and also a variable step P-stable method
of order 6 and 8 with a phase lag of the same order due to Simos
\cite{simos}. For the embedded Runge-Kutta, formulae of order 6(5)
and 8(7) have been developed by Prince and Dormand \cite{prince}
and formulae of order 8(6) and 12(10) have been developed by
Dormand {\it et. al.} \cite{dormand}. These integration algorithms
should be more efficient for higher accuracy. Our choice of the
integration algorithm is based on the following reasons. First,
our main objective has been to find a solution to the problem of
moving singularities and not the efficiency of the integration
algorithm itself. Second, it turns out that in this problem the
cumulative error of even the fourth order Runge-Kutta integration
algorithm is negligible compared to the error introduced to the
solutions at each jump across the singularities.

The set of equations that we discuss results from a classical
model of gravitation in Robertson-Walker cosmology in which the
signature of the metric  undergoes a transition from a Euclidean
to a Lorentzian domain. In section 2 we briefly discuss the
physical origin of the problem and show its reduction to a set of
ordinary differential equations. There, one sees an example in
which this  set  automatically includes a subsidiary equation
which is an integral of motion, along with the dynamical
equations. It is worth mentioning that if an integral of motion is
not directly included in the set of equations of motion, it can
some times be derived directly from the dynamical equations. Also
we employ a reparameterization transformation which allows one to
seek continuous solutions across the hypersurface of signature
change. Moreover, we employ a set of transformations which reduces
the degree of severity of the moving singularities. The reader who
is interested only in the numerical methods can skip to section 3,
without loss of continuity.

\section{Derivation of dynamical equations}

Traditionally, one of the features of classical gravity is that
the signature of the metric is usually considered as fixed. If one
relaxes this condition, one may find solutions to the field
equations which exhibit a signature transition \cite{Dereli,us}.
In the model that we study here a real scalar field is taken as
the matter source interacting with gravity and itself in a
Robertson-Walker geometry whose signature evolution is controlled
by a preferred coordinate. In this model, we seek solutions to the
dynamical equations which are smooth and continuous across the
hypersurface of signature transition, where the metric is
degenerate. The alternative would have been to find solutions by
solving Einstein's equations in disjoint regions next to the
hypersurface, and then finding jump conditions to match them
\cite{hel}. For the spatially flat universes, the first approach
yields  exactly solvable Einstein's equations \cite{Dereli}. Here,
we discuss the general case which includes the spatially flat as
well as non-flat cases and solve the resulting dynamical equations
numerically. For more details of the physical basis and
significance of the problem, we refer the interested reader to the
reference \cite{us}.

Consider gravity coupling to a scalar field through Einstein's
equation,
\begin{eqnarray}
G_{\mu\nu}=\kappa T_{\mu\nu} [\phi],  \label{eq1}
\end{eqnarray}
where the scalar field $\phi$ is a solution of the Klein-Gordon
equation,
\begin{eqnarray}
\Delta\phi-\frac{\partial U}{\partial\phi}=0. \label{eq2}
\end{eqnarray}
Here, $G_{\mu\nu}$ is the Einstein tensor constructed from
torsion-free connections compatible with the metric, and $U(\phi)$
is the scalar potential for the real scalar field $\phi$, which
interacts with itself and gravity through the stress-energy tensor
$T[\phi]$.

The above coupled equations are to be solved in a domain that
would lead to Robertson-Walker cosmologies with Lorentzian
signature. However, if the metric is suitably parametrized, one
expects to see continuous transition to a Euclidean domain. As in
\cite{Dereli}, we adopt a chart with coordinate functions
$\{\beta,x^1,x^2,x^3\}$ where the hypersurface of signature change
would be located at $\beta=0$. The metric can be parametrized to
take the form
\begin{eqnarray}
g=-\beta d\beta\otimes d\beta+\frac{R^2(\beta)}{[1+(k/4)r^2]^2}
\sum_i dx^i\otimes dx^i, \label{eq4}
\end{eqnarray}
where $r^2=\sum_i x^i x^i$. We seek solutions of the form
$R=R(\beta)$ and $\phi=\phi(\beta)$. Now, it is apparent that the
sign of $\beta$ determines the geometry, being Lorentzian if
$\beta>0$ and Euclidean if $\beta<0$. For $\beta>0$, the
traditional cosmic time can be recovered by the substitution
$t=(2/3)\beta^{3/2}$. Adopting the chart $\{t,x^i\}$ and using
equations (\ref{eq1}) through (\ref{eq4}) with units in which
$\kappa=1$, one finds
\begin{eqnarray}
-3\frac{\dot{R}^2}{R^2}-3\frac{k}{R^2}&+&
\frac{\dot{\phi}^2}{2}+U(\phi)=0, \label{eq5}\\
2\frac{\ddot{R}}{R}+\frac{\dot{R}^2}{R^2}+\frac{k}{R^2}&=&
-\frac{\dot{\phi}^2}{2}+U(\phi), \label{eq6}\\
\ddot{\phi}+3\frac{\dot{R}}{R}\dot{\phi}+\frac{\partial U}{\partial\phi}&=&
0, \label{eq7}
\end{eqnarray}
where a dot represents differentiation with respect to $t$ and
$-\infty<\phi<\infty$, $0\le R<\infty$.  Now a solution to the
problem is furnished by finding $R(t)$ and $\phi(t)$, for a given
$U(\phi)$. Note that these equations are not all independent. For
example equation (\ref{eq6}) can be obtained by combining
equations (\ref{eq5}) and (\ref{eq7}). Upon a closer inspection we
recognize that this is due to the fact that equation (\ref{eq5})
is not a dynamical equation, rather it is actually an integral of
motion representing a zero energy condition. That is, any solution
of the dynamical equations (\ref{eq6}) and (\ref{eq7}) would yield
a constant total energy (equation (\ref{eq5})). However,
Einstein's equations demand zero energy solutions only.

As is apparent from the dynamical equations, we have moving
singularities at all times for which $R=0$. These moving
singularities are potentially very severe and, as we shall see
later, $\phi$ actually diverges there. We can get an indication on
the divergence of $\phi$ from equations (\ref{eq5},\ref{eq6}).
These equations indicate that $U(\phi)$ has to cancel the
divergence of the $k/R^2$ terms, and for all physically relevant
potentials, this implies that $\phi$ has to diverge. We therefore
need to use a set of transformations to reduce the severity of the
divergence of solutions. We expect the following transformations
to render the solutions more manageable, since it is formed of
products of factors which go to zero and infinity at about the
same strength,
\begin{eqnarray}
X=R^{3/2}\cosh(\alpha\phi), \label{eq8}\\
Y=R^{3/2}\sinh(\alpha\phi), \label{eq9}
\end{eqnarray}
where $\alpha^2=\frac{3}{8}$.

The above equations are considerably simplified if we take the potential to be
\begin{eqnarray}
2\alpha^2(X^2-Y^2)U(\phi(X,Y))=a_1X^2+a_2Y^2+2bXY, \label{eq99}
\end{eqnarray}
where $a_1$, $a_2$ and $b$ are adjustable parameters. This choice for
the potential stems from the fact that the left hand side of equation
(\ref{eq99}) directly appears in the Lagrangian, from which the
dynamical eqations can be derived.
The features of this
potential and the physics involved in the choice of its parameters
have been discussed in \cite{Dereli,us}.

The dynamical equations (\ref{eq6}) and (\ref{eq7}) in terms of
$X$ and $Y$ and the evolution variable $\beta$ now become,
\begin{eqnarray}
Y^{\prime\prime}=\frac{1}{2}\left(\frac{1}{\beta}\right)Y^\prime-
\frac{3}{4}\beta kY(X^2-Y^2)^{-2/3}-\beta(a_2Y+bX), \label{eq10}\\
X^{\prime\prime}=\frac{1}{2}\left(\frac{1}{\beta}\right)X^\prime-
\frac{3}{4}\beta kX(X^2-Y^2)^{-2/3}+\beta(a_1X+bY), \label{eq11}
\end{eqnarray}
subject to the subsidary
`zero energy condition', equation (\ref{eq5}), which can be
written in terms of the new variables as
\begin{eqnarray}
\left(\frac{1}{\beta}\right)(-X^{\prime^{\,2}}+Y^{\prime^{\,2}})-
\frac{9}{4}k(X^2-Y^2)^{1/3}+(a_1X^2+a_2Y^2+2bXY)=0. \label{eq12}
\end{eqnarray}
Here, a prime represents differentiation with respect to $\beta$.
Note that this equation does not contain any singularity, and
equations (\ref{eq10}) and (\ref{eq11}) are actually less singular
than equations (\ref{eq6}) and (\ref{eq7}). The coupled equations
(\ref{eq10}) and (\ref{eq11}) must now be solved and, as explained
before, equation (\ref{eq12}) is merely a restriction on the
initial conditions. However, it can also be used as a consistency
check on the analytical or numerical solutions. These equations do
not seem to have a closed form solution so  a numerical treatment
is necessary.

\section{The numerical method}

The dynamical equations that we have to solve are equations
(\ref{eq10}, \ref{eq11}). As mentioned in the last section,
equation (\ref{eq12}) is an integral of motion. That is, any true
solution to equations (\ref{eq10}, \ref{eq11}) automatically
satisfies equation (\ref{eq12}) for all values of the
independent parameter, if it satisfies it at any one point.
Therefore, we can use equation (\ref{eq12})
to put a restriction on the initial conditions.
More importantly, one can check
the accuracy of the solutions by seeing how well
equation (\ref{eq12}) is satisfied,
as the algorithm integrates the dynamical equations.

As a first step towards a numerical solution to the equations, we
should study the restrictions imposed by the set of differential
equations on the initial conditions. These restrictions are the
result of the requirement of consistency of the initial conditions
with the dynamical equations. However we can accomplish a more
complete task by finding the general form of the analytic
solutions close to the initial point. These solutions certainly
include the complete information on the allowed set of the initial
conditions\footnote{The question of the allowed set of the initial
conditions, though interesting enough in its own right in all
problems of this type, is of crucial importance for the problem at
hand, as the determination of the correct initial conditions is an
open problem in cosmology.}. Moreover, the knowledge on the
analytic solutions help with the first few steps of the
integration algorithm.

\subsection{Analytic solutions close to the initial point}

In order to find analytic solutions which are valid near the
initial point ($\beta =0$), we first study the restrictions
imposed by  equations (\ref{eq10})--(\ref{eq12}) on the initial
conditions. This is done by noting that in order to have well
behaved  solutions close to $\beta=0$, the first term of equation
(\ref{eq12}) shows that  we must either have $X^\prime(\beta)\sim
\beta^{n_x}$ and $Y^\prime(\beta)\sim \beta^{n_y}$, where $n_x$,
$n_y\ge 1/2$, or $|X^ \prime(0)|=|Y^\prime(0)|$. However, the
first terms on the right hand sides of equations (\ref{eq10}) and
(\ref{eq11})  impose a more severe restriction. These two
equations admit solutions $X^\prime(\beta)\sim \beta^{1/2}$ and
$Y^\prime (\beta)\sim \beta^{1/2}$ close to $\beta=0$, however,
this class of solutions does not admit real or $C^2$ solutions
across $\beta=0$. One can show that regular solutions close to
$\beta=0$ are of the form
\begin{eqnarray}
X(\beta) &=& A_x \beta^3+X_0, \hspace{5mm} \mbox{where}
\hspace{5mm}  A_x= \frac{2}{9}
\left[-\frac{3}{4}\frac{kX_0}{(X^2_0-Y^2_0)^{2/3}}+a_1
X_0+bY_0\right], \label{eqxc0}\\
Y(\beta) &=& A_y \beta^3+Y_0,
\hspace{5mm} \mbox{where} \hspace{5mm}  A_y= \frac{2}{9}
\left[-\frac{3}{4}\frac{kY_0}{(X^2_0-Y^2_0)^{2/3}}-a_2
Y_0-bX_0\right], \label{eqyc0}
\end{eqnarray}
with $X_0\equiv X(0)$, etc. Therefore,  the initial conditions on
the first and second derivatives must satisfy the relations
\begin{eqnarray}
X^\prime(0)=Y^\prime(0)=0\hspace{3mm} \mbox{and}\hspace{3mm}X^{\prime\prime}
(0)=Y^{\prime\prime}(0)=0.\label{eq15}
\end{eqnarray}

\begin{figure}[t,b]
\vspace{-2.5cm} \centerline{\begin{tabular}{cc}
\hbox{\epsfxsize=6.9cm\epsfysize=7.0cm\epsffile{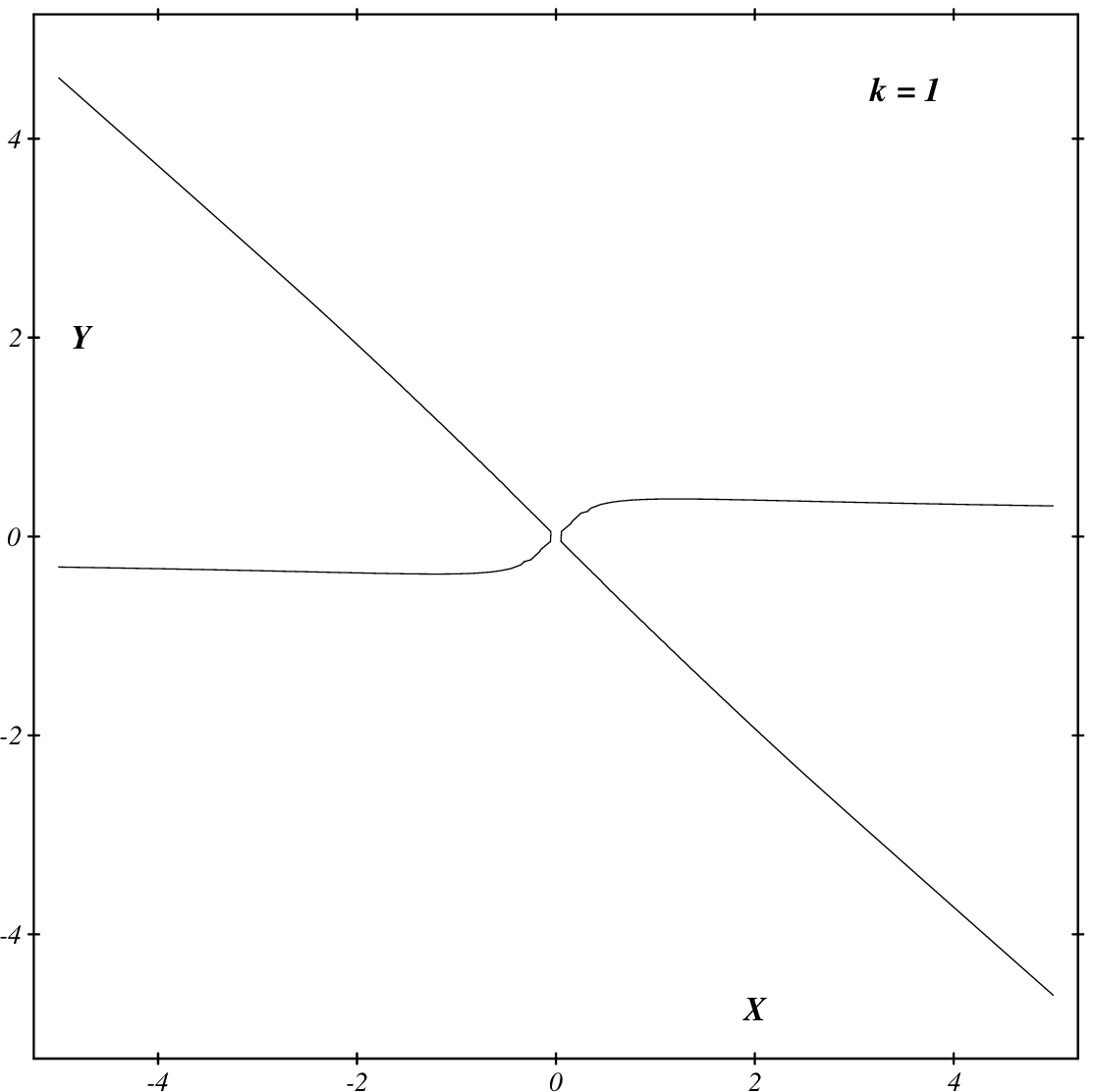}}&\hspace{1cm}
\hbox{\epsfxsize=6.9cm\epsfysize=7.0cm\epsffile{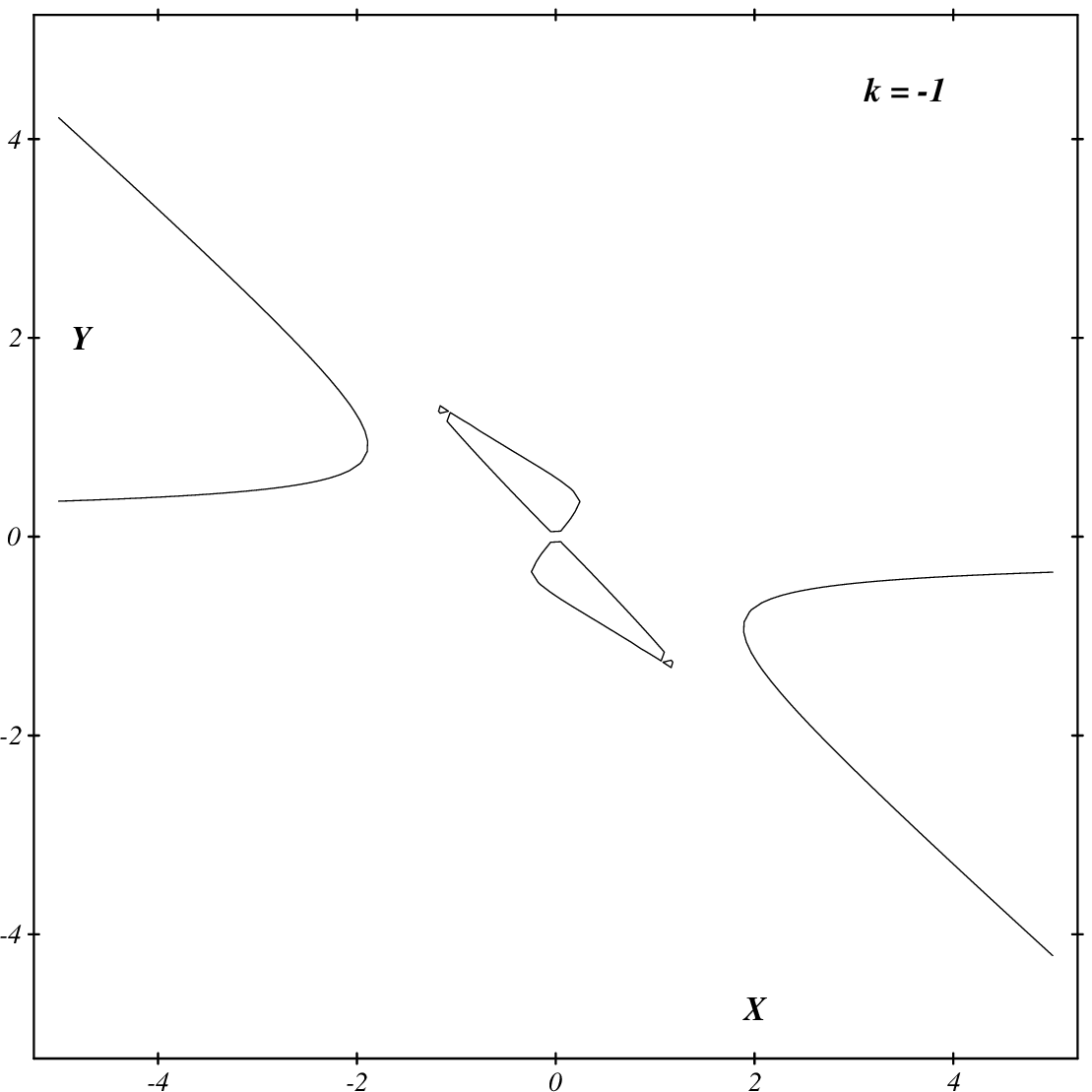}}\end{tabular}}
\vspace{0.5cm} \caption{\footnotesize The contour plots of the
allowed initial values of $X$ and $Y$, satisfying the equation of
constraint (\ref{eq16}) for $k=\pm1$. The point (0,0) is a
solution and the curves approaching this point actually pass
through it, although this is not shown on the plots due to the
limitations on the numerical accuracy.} \label{fig1}
\end{figure}

Strictly speaking the conditions on the second derivatives are not
initial conditions but rather consistency checks, since we have
coupled second order equations. Therefore, the initial values for
the functions $X$ and $Y$ must now satisfy, c.f. equation
(\ref{eq12}),
\begin{eqnarray}
-\frac{9}{4}k(X^2_0-Y^2_0)^{1/3}+(a_1X^2_0+a_2Y^2_0+2bX_0Y_0)=0. \label{eq16}
\end{eqnarray}
The contour plots of equation (\ref{eq16}) for $k=\pm1$ are  given
in figure \ref{fig1}. Along the contours, one finds the possible
initial values for $X$ and $Y$. Although equation (\ref{eq16}) is
equivalent to a sixth order algebraic equation which cannot be
directly solved  analytically, we can solve it by going back to
the original variables $R$ and $\phi$. The solutions are either
$R(0)=0$ giving $\phi(0)=\pm\infty$, which we exclude because we
have been seeking continuous solutions across $\beta=0$, or
$R(0)\ne 0$ (it is a free parameter) with
\begin{eqnarray}
\phi(0)=\frac{1}{2\alpha}\cosh^{-1}\left[\frac{DB\pm b\sqrt{D^2-B^2+b^2}}{B^2-
b^2}\right],
\label{eq166}
\end{eqnarray}
where
\begin{eqnarray}
D=\frac{9k}{4R(0)^2}-\frac{a_1-a_2}{2}\hspace{5mm}\mbox{and}\hspace{5mm}B=
\frac{a_1+a_2}{2}=\frac{m^2}{2}.\nonumber
\end{eqnarray}
Therefore, the acceptable values of $X_0$ and $Y_0$ ( $X_0>|Y_0|$) can also be
obtained analytically from equation (\ref{eq166}).

\subsection{Integration algorithm}

The important feature of equations (\ref{eq10}) and (\ref{eq11})
is that they are singular for all $\beta$ at which $X=\pm Y$.
At these critical
values of $\beta$  ($\beta_c$), the original variables take the values
$R(\beta_c)=0$ and $\phi(\beta_c)=\pm\infty$. We can directly infer
from the differential equations that the solutions for the new variables
and their first
derivatives have to be continuous across the singularities. However,
the second and higher derivatives will be singular at $\beta_c$. That is,
the singularities of the new variables are considerably
milder than those of the original variables.

Although the solutions and their first derivatives are continuous
across $\beta_c$, they cause problems for the integration algorithm.
Any attempt in
solving these equations involves handling these moving
singularities, as one encounters them when integrating the
coupled equations. To proceed, we first establish
jump conditions across these singular points as follows: close
to $\beta_c$ we assume that the solutions have the
following linear forms
\begin{eqnarray}
X_{\pm}& = & a_{\pm}+b_\pm\beta, \label{lin1} \\ Y_{\pm}& = &
c_\pm+d_\pm\beta, \label{lin2}
\end{eqnarray}
where $\pm$ refers to the right or left hand sides of the
singularity, respectively. Substituting the above equations in
(\ref{eq10}) and (\ref{eq11}) and dropping all non-singular terms,
one can integrate these equations in the interval
$\beta_c-\epsilon,\hspace{1mm}\beta_c+\epsilon$, where $2\epsilon$
is the distance across the jump. For the integration we have
dropped all terms which would give rise to contributions ${\cal
O}(\epsilon^{4/3})$. One finds at $ Y_c=\pm X_c$
\begin{eqnarray}
b_{+}-b_{-}=-\frac{9}{4}k\frac{(2X_c\epsilon)^{1/3}}{(b_{-}\mp
d_{-})^{2/3}}\beta_c=\mp(d_{+}-d_{-}), \label{eq155}
\end{eqnarray}
where $\epsilon$ can be taken as small a value as is desired for
any required accuracy. Equation (\ref{eq155}), together with the
requirement of continuity of $X$ and $Y$, establish our jump
condition for handling the singularities of the differential
equations. It is apparent from equation (\ref{eq155}) that the
slopes $X^\prime (\beta)$ and $Y^\prime (\beta)$ are continuous at
$\beta_c$.

Writing an actual  algorithm for handling these singularities
requires some care. Let us first write the original variable
$R(\beta)$ in terms of the new variables, $$ R=(X^2-Y^2)^{1/3}.$$
We recall that the differential equations become singular when
$R=0$. As we integrate them, when $|R|$ becomes small (less than
1) we reduce the step size by one order of magnitude since the
crossing of $R(\beta)$ through zero at $\beta_c$ is rather steep.
Then at the first instant when $|R|$ becomes smaller than 0.1,
henceforth called `the fixed point', the algorithm  records all
the relevant values ($X,Y,X^\prime,Y^\prime,R,\beta$) and
continues integrating towards the singular point with yet finer
steps. Past the fixed point, the singular terms in the
differential equations become too large and no integration
algorithm can give reliable  values for $X$ and $Y$. However, we
can use the information obtained past this point to pinpoint
$\beta_c$ as follows: We record the last two values of $\beta$ and
$R$ right before the instant when the sign of $R$ changes. Then
assuming $R\propto (\beta-\beta_c)^{1/3}$ (which is consistent
with equations (\ref{lin1}) and (\ref{lin2})) one obtains
\begin{equation}
\beta_c=\frac{R_2^3\beta_1-R_1^3\beta_2}{R_2^3-R_1^3}.\label{bet}
\end{equation}
Having obtained $\beta_c$, we can calculate
$\epsilon=\beta_{\mbox{\scriptsize fixed}}-\beta_c$ and use a
linear extrapolation to obtain $X_c$ and $Y_c$ and see whether
$Y_c=\pm X_c$ as a consistency check. We can then calculate the
values of the slopes ($X^\prime,Y^\prime$) on the other side of
the singularity (at $\beta_{\mbox{\scriptsize
over}}=\beta_c+\epsilon$) using the jump conditions (equation
(\ref{eq155})) and then using linear extrapolation on both sides
of the singularity, the values of the functions can be calculated
at $\beta_{\mbox{\scriptsize over}}$. The algorithm then continues
integrating with fine steps until the value of $|R|$ increases
beyond 1 and with regular steps until it approaches the next
singularity.

\begin{figure}
\vspace{-2.5cm} \centerline{\begin{tabular}{cc}
\hbox{\epsfxsize=6.9cm\epsfysize=7.0cm\epsffile{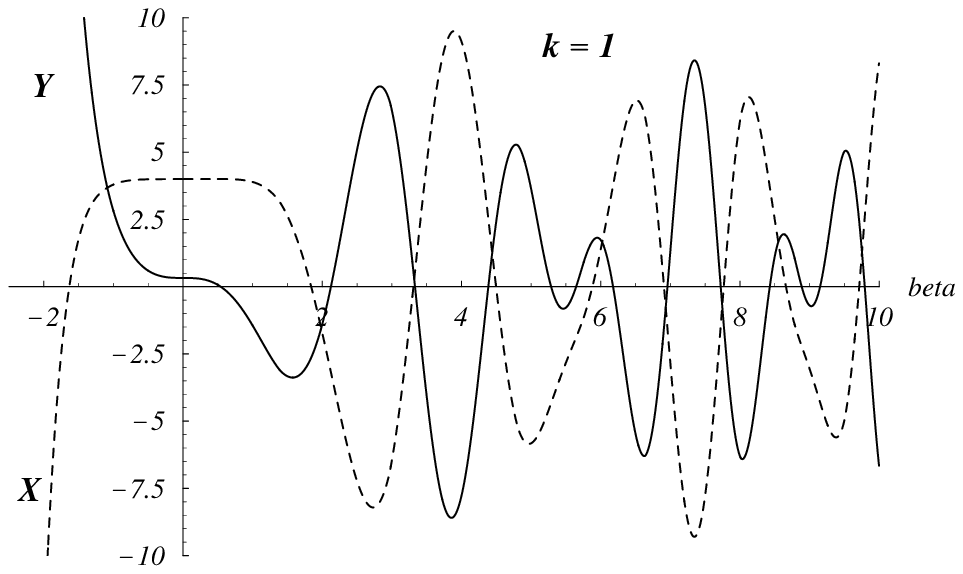}}&\hspace{1cm}
\hbox{\epsfxsize=6.9cm\epsfysize=7.0cm\epsffile{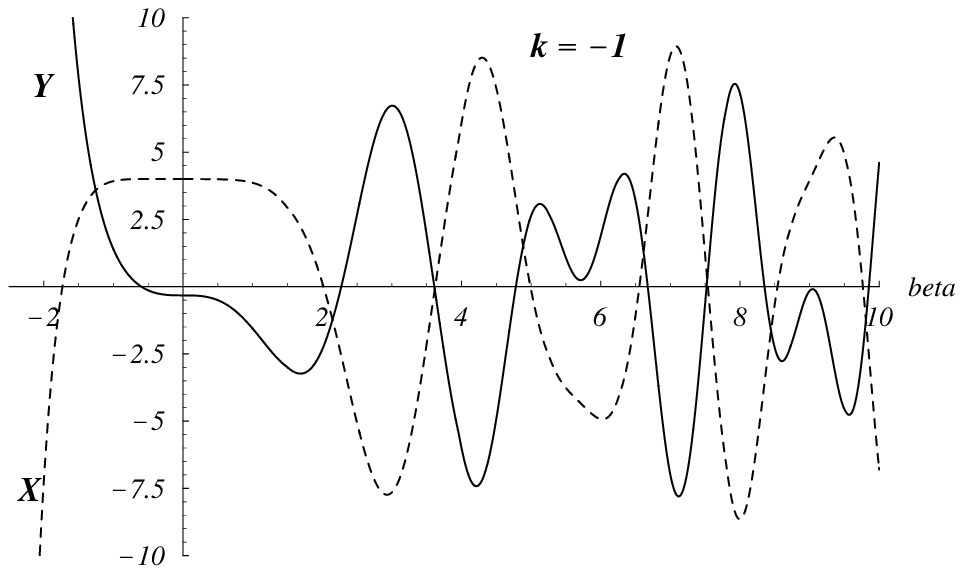}}\end{tabular}}
\caption{\footnotesize Solutions for $X(\beta)$
(broken curve) and $Y(\beta)$ (solid curve), for $k=\pm1$. The
values of the parameters are $b=2$, $\lambda=0$, $m^2=4.5$.}
\label{fig2}
\end{figure}

\begin{figure}
\vspace{-2.5cm} \centerline{\begin{tabular}{cc}
\hbox{\epsfxsize=6.9cm\epsfysize=7.0cm\epsffile{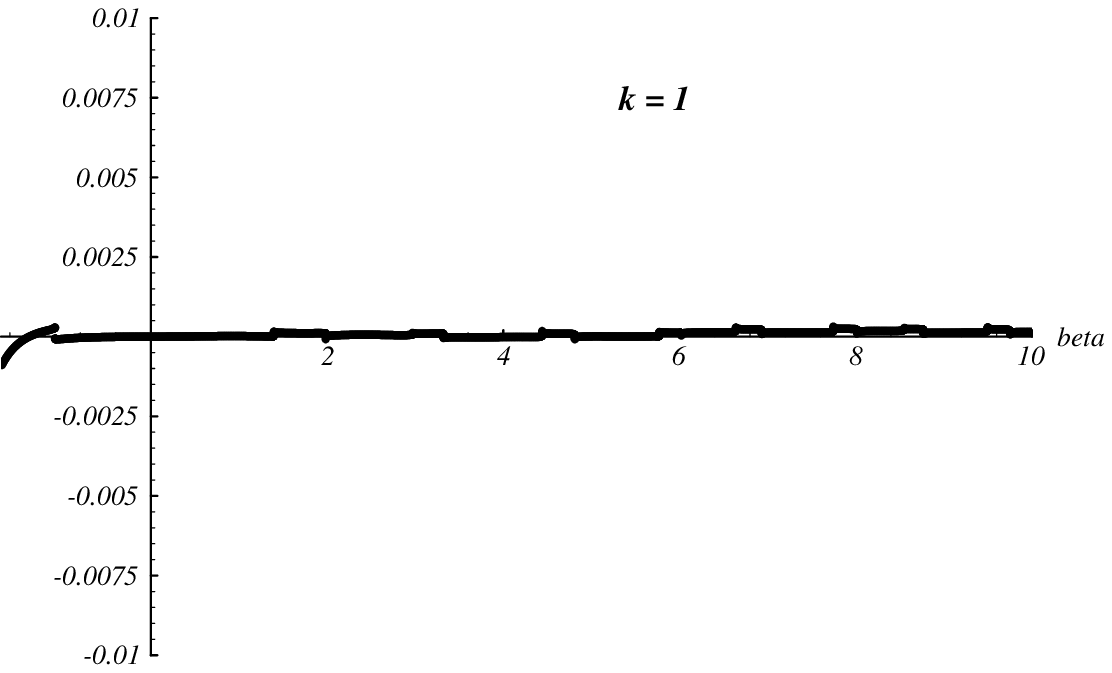}}&\hspace{1cm}
\hbox{\epsfxsize=6.9cm\epsfysize=7.0cm\epsffile{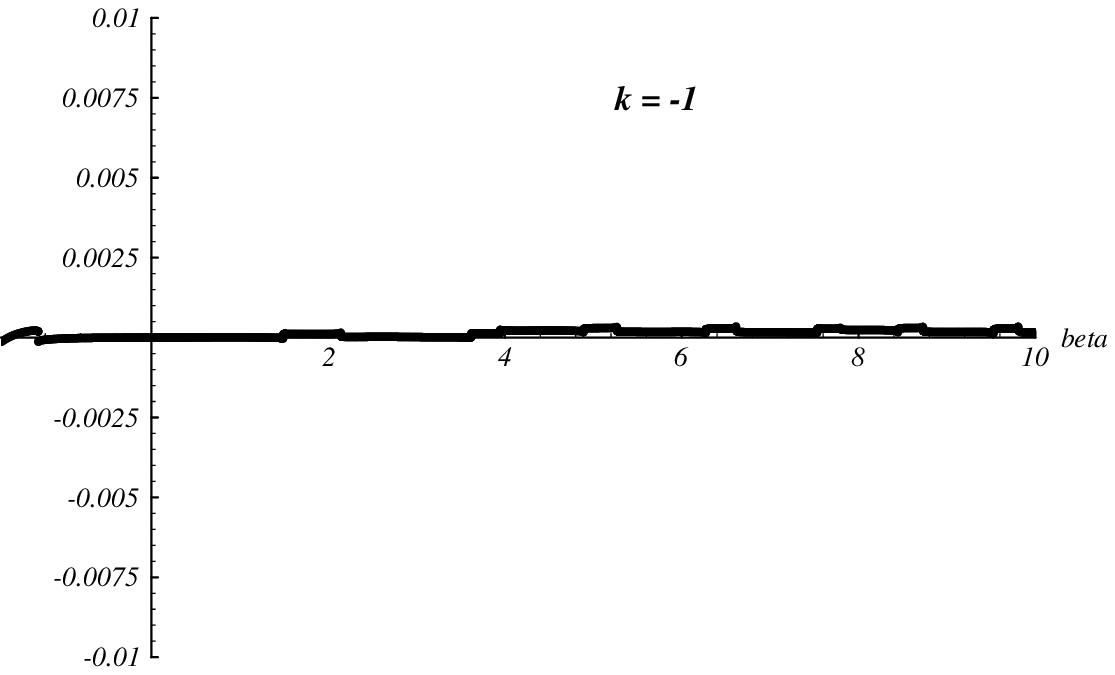}}\end{tabular}}
\caption{\footnotesize The graph of the total
energy defined by equation (\ref{eq12}) for $k=\pm1$. As is
apparent from the graphs, the zero energy condition is satisfied
to a high accuracy. The small jumps in the graphs are at the
critical values of $\beta$ where there are singular points:
$Y(\beta_c)=\pm X(\beta_c)$.} \label{fig3}
\end{figure}

\begin{figure}
\vspace{-1.5cm}
\centerline{ 
\hbox{\epsfxsize=10.0cm\epsfysize=7.0cm\epsffile{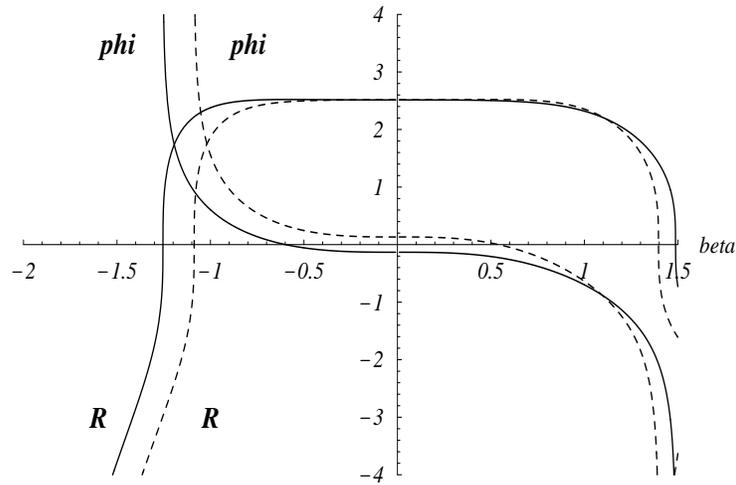}}}
\caption{\footnotesize Graphical representation of
the original variables $\phi(\beta)$ and $R(\beta)$, for $b=2$,
$\lambda=0$, $m^2=4.5$, and $k=1$ (broken curves) and $k=-1$
(solid curves).} \label{fig4}
\end{figure}

We use a set of parameters ($b=2$, $\lambda=0$, $m^2=4.5$) in
equations (\ref{eq10}-\ref{eq12}) which are physically relevant
and choose our initial conditions consistent with equations
(\ref{eq15}) and (\ref{eq16}). Recall that since equation
(\ref{eq12}) is a constant of motion, if it is satisfied at
$\beta=0$, for a true solution it will be satisfied at all other
values of $\beta$. Therefore if the initial values for the
functions $X$ and $Y$ satisfy equation (\ref{eq16}) at $\beta=0$,
equation (\ref{eq12}) should always be satisfied. For integrating
equations (\ref{eq10}) and (\ref{eq11}), we have used the fourth
order Runge-Kutta method. The resulting solutions for $k=\pm 1$
are shown in figure \ref{fig2}.  It is apparent from figure
\ref{fig2} that at the singular points $X=\pm Y$, the solutions
are continuous and the singularities are very mild. As a measure
of the accuracy of the solutions, we have computed the `zero
energy condition', equation (\ref{eq12}), as a function of $\beta$
for $k=\pm1$ which are shown in figure \ref{fig3}. It is evident
from figure \ref{fig3} that the values of `total energy' stay very
close to zero, thus indicating the validity of the numerical
solution. In figure \ref{fig4} the variations of the original
variables $\phi$ and $R$ are shown. As can be seen, $\phi$
actually diverges at the singular points. As a further check, we
have numerically recovered the analytic solutions presented in
\cite{Dereli} for $k=0$ in every detail.

\section{Conclusions}

We have shown how a particular class of initial valued coupled
non-linear ordinary differential equations with moving
singularities can be numerically solved. The main obstacle of having
moving singularities can be overcome by establishing a set of
jump conditions across them. These conditions are obtained by
approximating the form of the solutions close to the singular
points and directly integrating the differential equations in
the neighbourhood of these points. We have found that a first
order approximation close to these points is sufficiently
accurate. Also since the main source of error in the solutions
eminates from the jump conditions, we have found that the fourth
order Runge-Kutta is sufficient for the integration algorithm.


\end{document}